\documentclass[preprint2]{aastex}

\newcommand{\chg}[1]{#1}

\usepackage{psfig}

\makeatletter

\newenvironment{inlinefigure}{%
\def\@captype{figure}%
\noindent\begin{minipage}{0.999\linewidth}\begin{center}}
{\end{center}\end{minipage}\smallskip}
\makeatother

\slugcomment{Accepted to the Astronomical Journal}

\shortauthors{Miyaji et al.}
\shorttitle{Chandra Observations of z$\sim 3$ QSOs}

\begin{document}


\title{Chandra Observations of Six QSOs at z$\approx$ 3
 \footnote{Based on observations using the Chandra X-ray Observatory.}}


\author{Takamitsu Miyaji}
\affil{Physics Department, Carnegie Mellon University,
     5000 Forbes Ave., Pittsburgh, PA 15213}
\email{miyaji@cmu.edu}
\author{G\"unther Hasinger, Ingo Lehmann\altaffilmark{1}}
\affil{Max-Planck Institut f\"ur extraterrestrische Physik,
Postf. 1312, 85741 Garching, Germany}
\author{Donald P. Schneider}
\affil{Department of Astronomy and Astrophysics, 525 Davey Laboratory,
Pennsylvania State University, University Park, PA 16802}

\altaffiltext{1}{Present Address:LPS-Berlin, K\"opernickerstr. 325, 12555, Berlin, 
Germany}

\begin{abstract}
 We report the results of our {\it Chandra} observations of six QSOs
at $z\sim 3$ from the Palomer Transit Grism Survey. Our primary goal
is to investigate the possible systematic change of $\alpha_{\rm ox}$
between $z>4$ and $z\sim 3$, between which a rapid rise of luminous
QSO number density with cosmic time is observed. The summed spectrum
showed a power-law spectrum with photon index of $\Gamma \approx 1.9$,
which is similar to other unabsorbed AGNs. Combining our $z\sim 3$ 
QSOs with X-ray observations of QSOs at $z>4$ from literaure/archive, 
we find a correlation of  $\alpha_{\rm ox}$ with optical luminosity.
\chg{This is consistent with the fact that the luminosity 
function slope of the luminous end of the X-ray selected QSOs is steeper
than that of optically-selected QSOs.}  We discuss an upper 
limit to the redshift dependence of  $\alpha_{\rm ox}$ using a 
Monte-Carlo simulation. Within the current statistical errors including 
\chg{the derived limits on the redshift dependence of $\alpha_{\rm ox}$}, 
we found that the behaviors of the X-ray and optically-selected 
QSO number densities are consistent with each other.  
\end{abstract}


\section{Introduction}
\label{sec:intr}

 Large-scale optical surveys show that the
luminous QSO number density peaks at $1.7\le z\le 2.7$, before
which (in cosmic time) the QSO number density grows rapidly
and after which the density steadily decays until
the present epoch \citep[e.g.][]{boy88,who94,ssg95}. 
The first indication of a density decline (with redshift) at $z\ga 3$
was reported in the pioneering work of \citet{osm82}.
The Palomar Transit Grism Survey (PTGS) \citep{ssg94} was designed 
to investigate this
possible ``redshift cutoff", and produced a sample of 90 $z>2.7$
QSOs; an analysis of the PTGS \citet{ssg95} (hereafter SSG95) revealed a very
rapid growth, by a factor of $\sim 3-5$, of the number density of
luminous QSOs from $z\sim 4.5$ to $\sim 2.7$.  This result was
verified by \cite{fan01} using a sample of $3.6<z<5.0$ quasars found
in the Sloan Digital Sky Survey \citep[SDSS;][]{sdss}. This redshift
cutoff has also been found in the number densities of radio flat-spectrum 
QSOs \citep{shav96,wall05}.

In the X-ray regime, however, the size
of a sample based on the combination of various {\it ROSAT} surveys,
\citet{xlf1} (heraafter, SXLF1) was not sufficient to accurately
probe this ``growth'' phase of the number density of X-ray-selected
luminous ($\log\,L_{\rm x}>44.5$) QSOs. While this study appears
to show a flat number density in $z>2.7$, the uncertainties produced by
the small number of quasars limited the robustness of the conclusions.
More recent studies including the results from {\it Chandra} 
and {\it XMM-Newton} surveys \citep{cowie03,fiore03,ueda03,barger05} 
showed that the number density curve is luminosity-dependent and a 
trend that as luminosity goes lower, the density peak shifts to 
lower redshifts
(as some authors call an  anti-hierarchical AGN evolution and others
call a down-sizing of the AGN activity). 
However, these studies did not sample well the high-redshift, 
high-luminosity regime, in which optically selected QSOs  
show a decline at $z>3$.

 In a recent study based on an updated soft X-ray sample including 
{\it Chandra} and {\it XMM-Newton} surveys, \citet{sxlf3} 
(hereafter SXLF3) also measured the number densities in various 
luminosity bins with a better accuracy over a large range in the 
redshift-luminosity space. The study revealed this early ``growth'' 
of AGN number density (or decline with increasing $z$) at 
z$\gtrsim 3$ for AGN/QSO luminosities at 
($\log\,L_{\rm x}\la$ 44-45). The results from ``Champs'' survey,
which was designed to optimally trace the high-redshift, 
high-luminosity ($\log L_{\rm x}>44.5$) regime with improved statistics, 
found that the density curves X-ray selected QSOs ($\log L_{\rm x}>44.5$) 
declines with increasing redshift at $z>3$. This decline is, however, 
shallower than that seen in optically-selected QSOs.

  The dependence of the optical to X-ray flux ratio (customarily
expressed by the quantity $\alpha_{\rm ox}$, the effective
spectral index between the rest-frame
2500 \AA\, and 2 keV)
\footnote{We use $f_\nu \propto \nu^{\alpha_{XX}}$, where XX is 
any subscript to $\alpha$.}
on redshift and luminosity has been a key issue in X-ray observations of
high-redshift QSOs and has important implications for possible
differences in the AGN evolution traced by X-ray and optical samples.
Some authors \citep[e.g.][]{vig03b,str05}
found that $\alpha_{\rm ox}$ strongly depends
on luminosity, with $L_{\rm x}\propto L_{\rm opt}^{0.75}$, and no evidence
for any evolution of the X-ray properties with redshift.   
\citet{bech03} found that variations in $\alpha_{\rm ox}$ depends
primarily on redshift. The dependence may be sensitive to the selection
effects, including but not limited to whether the sample is optically-selected
or X-ray selected. \citet{yuan98b} pointed out that such an apparent 
non-linearity of the luminosity correlation in two bands can arise from 
the difference in the luminosity variations in the two bands.

In order to investigate the redshift dependence of the optical-to-X-ray
luminosity ratios and its impact on the density curves of luminous QSOs
in X-ray and optically selected samples at $z\gtrsim 3$,  
we have obtained {\it Chandra} observations of
six PTGS QSOs with redshifts between 2.91 and 2.96. This
is the era of maximum number density of luminous QSOs. There was 
practically no systematic observations in X-rays in this redshift regime
before. Thus our observations also serve to fill this observation gap.   
Throughout this paper we adopt
$(H_{\rm 0}{\rm [km\,s^{-1}\,Mpc^{-1}]},\Omega_m,\Omega_\Lambda)=
(70h_{70},0.3,0.7)$ and $h_{\rm 70}=1$ unless otherwise noted.

\section{Observations and Analysis}

\subsection{The Sample and Observation}

 The original motivation of the program was to compare the 
mean $\alpha_{\rm ox}$ values of $z\sim 3$ and $z\gtrsim 4$ QSOs.
In Chandra Cycle 4, six $z\sim 3$ QSOs from the PTGS (out of
15 QSOs proposed) have been observed. 
None of these QSOs are broad absorption line (BAL) QSOs. 
Table \ref{tab:log} shows the observed targets, log of observations, 
optical AB magnitudes at the object's rest frame of 1450 \AA.  
The core radio loudness  
$R_{\rm L}\equiv \log_{10} f_\nu [20cm]/f_\nu [4400$ \AA$]$
is also shown, where those with  $R_{\rm L}\leq 1$ and $>1$ 
are divided into radio-quiet QSOs (RRQ) and radio-loud QSOs (RLQ)
respectively (\citealt{wilkes} and references therein). The fluxes 
are in the object's rest frame, calculated
assuming radio and optical spectral indices of $\alpha_{\rm r}=-0.8$
and $\alpha_{\rm o}=-0.79$ respectively.
The radio data are from the NVSS \citep[PC 0041+0024][]{nvss}
or FIRST \citep[all others][]{first} surveys.  
Only one QSO (PC 1035+4747) was detected in the radio band and 
for others, 3$\sigma$ upper limits of $R_{\rm L}$ are shown. 
The only radio-detected QSO, PC 1035+4747, has $R_{\rm L}=2.3$; 
this object falls well into the RLQ regime. Two of the  
$R_{\rm L}$ upper limits (1.5 for PC 0041+0214 and 1.2 for
PC1000+4751) are above the RQQ/RLQ border, but their limits
are well below the peak of the $R_{\rm L}$ distribution of 
RLQ. Thus we tentatively classify them as RQQs.
 
 All observations have been made with the Advanced CCD 
Imaging Spectrometer (ACIS; \citealt{acis}) and the targets aimed at the 
default position of ACIS-S with the S3-chip.
In all cases except PC 0947+5628, X-ray counterparts have been found 
within $1.5\arcsec$ of the cataloged optical center of the target QSOs, 
consistent with the combined systematic error on the absolute 
astrometries  of 1$\arcsec$-2$\arcsec$ in both the {\it Chandra} data 
products and the PTGS survey.  Although the detected  X-ray source closest 
to PC0947+5628 was $2.8\arcsec$ away from the catalogued position of the 
QSO, 6 other X-ray sources in the same ACIS observation also had optical 
counterparts at  $\sim 2.8\arcsec$ away with practically the same offset
directions. Thus we also identify the X-ray source with  PC0947+5628. 

\placetable{tab:log}

\subsection{Individual and Stacked Spectra}

 For all six observations, we have extracted the pulse-height 
spectra using an extraction radius of 2$\arcsec$. 
Spectra, response matrices (rmf) and ancillary response files 
(arf) were created using the software package {\bf CIAO 3.0.2} or later
versions, in conjunction with the calibration database {\bf CALDB 2.26}
or later versions. These versions enabled construction of the response 
files which takes the time-dependent low energy efficiency degradation
into account. Due to low number counts of the involved
objects, changes due to further updates of the calibration
have negligible effects. The spectral analyses were made to the pulse-height 
channels corresponding to observed photon energies of 0.3-7 keV. 
Background level is typically $\sim 0.05$ counts in the extraction 
radius and is thus negligible. The spectral analysis were made with 
{\bf XSPEC 11.2}.  In spite of small number of X-ray 
photons, the negligible background and the use of the {\bf XSPEC} 
implementation of the \citet{cash} C-statistics allowed 
placement of some constraints on the spectral indexes, although there
are not sufficient number of photons in any individual spectrum to 
simultaneously constrain the intrinsic absorption column 
density. The results of the spectral fits with a 
single power-law with a photon index $\Gamma$ and the Galactic absorption 
$N_{\rm H20}^{\rm G}$ ${\rm [cm^{-2}]}$ \cite{nh} at the position 
of the QSOs are shown in Table \ref{tab:spec}.
We see that the only radio-detected QSOs photon index
of $\Gamma=0.4\pm 1.0$ is constrained to be harder than the 
mean QSO spectrum. The rest-frame 2-10 keV luminosities 
($\log\;L_{\rm x}$; logarithm is base-10) are also shown as well 
as the source counts in 0.3-7 keV.  The rest-frame 2-10 keV
corresponds to observed frame 0.5-2.5 keV. The B-band absolute 
magnitudes ($M_{\rm B}$), that have been 
recalculated using our default cosmology (Sect. \ref{sec:intr}) 
and the optical spectral index $\alpha_{\rm o}=-0.79$ 
\citep[following][]{vig03c,fan01}, are also listed here.  Using
$\alpha_{\rm o}=-0.5$ \citep{dps01} increases $M_{\rm B}$ by 
0.35.


\placetable{tab:spec}

\begin{inlinefigure}

\centerline{\includegraphics[width=\textwidth]{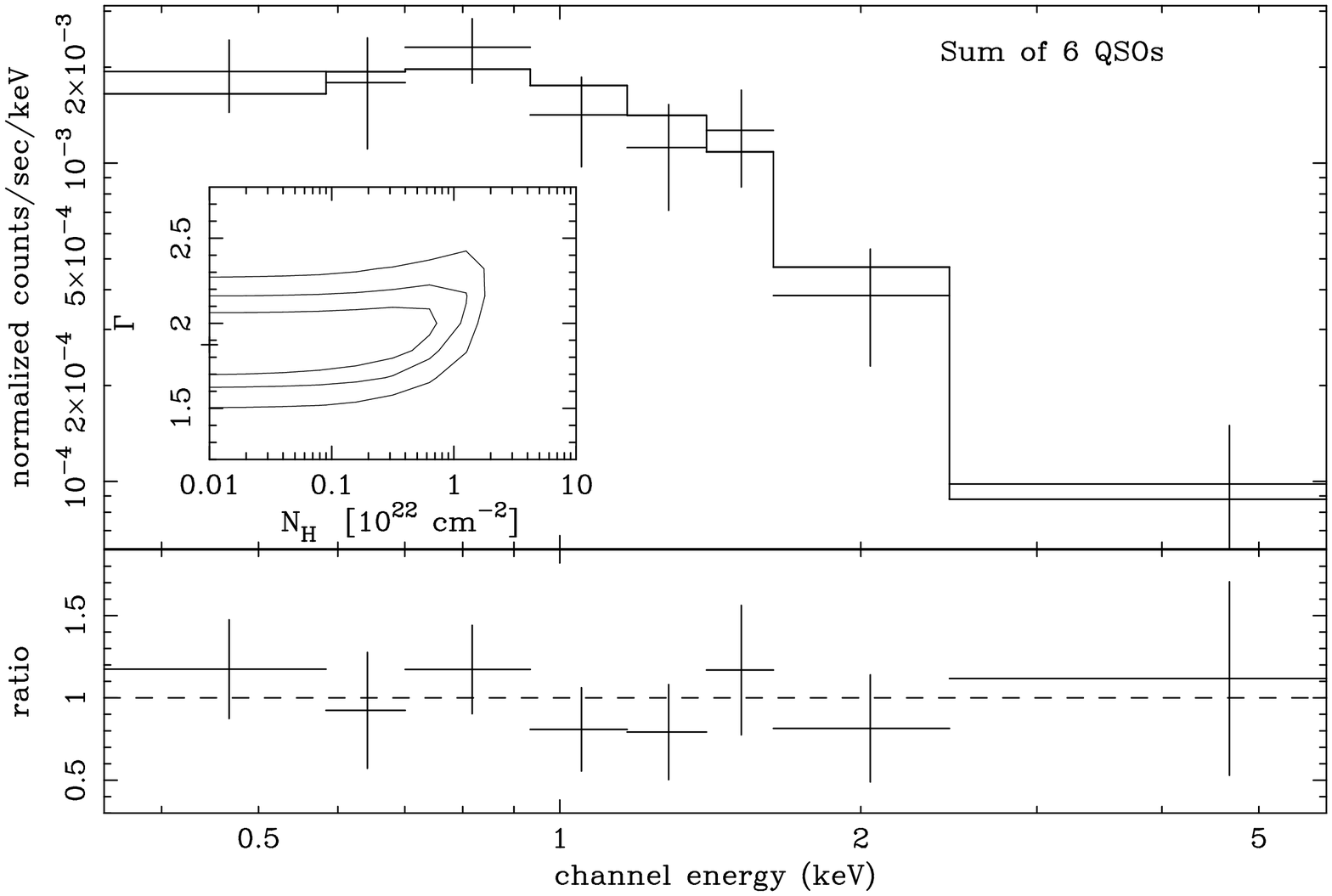}}
\caption
 {The stacked pulse-height spectrum of the 6 QSOs at $z\sim 3$ 
 with the folded best-fit power-law model and residuals
 in terms of the data-to-model ratio. The spectrum is rebinned for 
 display, but the actual fit was made to a higher resolution. 
 Confidence contours at levels of $\Delta C=2.3$ and 4.6, 9.2 
 (corresponding to the 68\%, 90\% and 99\% confidence levels for two 
 interesting parameters respectively) in the  
 $N_{\rm H22}-\Gamma$ plane are also shown.\label{fig:sum_spec}}
\end{inlinefigure}

\clearpage

 We also analyzed the summed spectrum from all six QSOs. 
{Because of the small spread of the redsifts of these QSOs, we can analyze
the summed spectrum assuming a single redshift.}  
The response matrix for the summed spectrum was constructed by
a source-count weighted mean the 6 matrices. 
\chg{The Galactic
column densities $N_{\rm H20}^{\rm G}$ of these six QSOs 
are 0.9, 1.0, 1.0, 1.3, 2.0 and 2.8. The fit was made
to a model with the sum of three power-laws, with different 
Galactic absorptions of $N_{\rm H20}^{\rm G}=1.0,2.0$ and 2.8 
respectively, where the four QSOs with $0.9\leq N_{\rm H20}^{\rm G}\leq 1.3$
were represented by a single column density of 1.0.} 
The photon indices of all the three components 
were set to equal and the ratios of the 3 normalizations were 
fixed to those of the total source counts of the QSOs with 
$N_{\rm H20}^{\rm G}$ of 0.9-1.3, 2.0 and 2.8 respectively. 
Also an intrinsic absorption component  $N_{\rm H20}^z$ is included with 
$z=2.93$, which is a source-count weighted mean redshift of the 
sample. Again, the C-statistics was used for the fit.
The summed spectrum is well represented by a single
power-law with $\Gamma \sim 1.9$ and no intrinsic absorption,
as shown in the last entry of Table \ref{tab:spec}. The pulse-height
spectrum, folded best-fit power-law model, and fit residuals in terms
of data-to-model ratio are shown in Fig. \ref{fig:sum_spec}, with 
confidence contours for the intrinsic absorption versus photon
index space. Removing the one RLQ  (PC 1035+4747) from the analysis 
did not change the fitted parameters and errors to the smallest digits 
displayed in Table \ref{tab:spec}.  This result is consistent with the 
mean slopes of RQQs and unabsorbed AGNs measured in the 2-10 keV in 
the rest frame over a wide range of redshift and luminosity 
\citep[e.g.][]{vig03a,vig03c}. Note, however, that the stacked spectrum is 
dominated by a few brightest sources, {with 60\% of the photons coming 
from the two brightest objects.} The quoted error only includes the 
statistical error of photon counts. 
The sampling error is estimated by a bootstrapping 
method, where the 90\% error range was determined by 500 bootstrap runs of 
a photon-count weighted mean best-fit $\Gamma$ values. The results were 
(90\% bootstrap errors) $\langle \Gamma \rangle=1.85^{+0.31}_{-0.25}$ 
($1.97\pm0.29$ with the RLQ removed). 
 
\subsection{The Optical to X-ray Index ($\alpha_{\rm ox}$)}

 The optical (rest-frame ultraviolet) to X-ray flux ratio of a QSO 
is customarily expressed in terms of the effective index 
$\alpha_{\rm ox}$ between 2 keV and 2500\AA\, in the QSOs rest frame.
Upon calculating $\alpha_{\rm ox}$, we assumed $\Gamma=2.0$ for all, 
which is the average QSO spectral index. 
The result of the spectral analysis of all but one is consistent 
with the canonical spectral index of $\Gamma=$ 1.9-2.0. 
The RLQ PC 1035+4747 has $\Gamma\sim 0.6\pm 0.8$ and using $\Gamma=0.6$
for K-correction gives $\alpha_{\rm ox}=-2.06$. Other than this one, 
the main source of errors in  $\alpha_{\rm ox}$ is the X-ray flux. 
Even for the source with the smallest source count (PC 1035+4747) 
the 1$\sigma$ error on $\alpha_{\rm ox}$ is $\sim 0.05$. A decrease of 
$\Gamma$ by 0.2 leads to an decrease of $\alpha_{\rm ox}$ 
by $0.04$ at $z\sim 3$.  Using $\alpha_{\rm o}=-0.5$
instead of $-0.79$, $\alpha_{\rm ox}$ increases by 0.03.  

\section{Redshift and $M_{\rm B}$ Dependences of $\alpha_{\rm ox}$}
 
 This program is mainly focused on the systematic difference 
in $\alpha_{\rm ox}$ between $z\sim 3$ and $z>4$.
Because $\alpha_{\rm ox}$ values of QSOs show a large scatter
and we only sample a small number of QSOs in the redshift-luminosity 
regimes of our interest, the sampling error is the dominant
effect in the error budget of the {\em mean value}
($\langle \alpha_{\rm ox} \rangle$) of QSOs, which can be
estimated by $\sigma/\sqrt{N_{\rm Q}}$, where $\sigma$ is the standard
deviation of the $\alpha_{\rm ox}$ distribution of $N_{\rm Q}$ QSOs.
\chg{Note that the $\sigma/\sqrt{N_{\rm Q}}$ estimation of the 
{\em standard deviation of the mean} is also valid for small $N_{\rm Q}$. 
It is well known that this estimator gives the exact confidence range 
of Gaussian 1$\sigma$ when the parent $\alpha_{\rm ox}$ 
distribution is a Gaussian and it is widely used in more general cases.
The $\alpha_{\rm ox}$ distributions in \citet[e.g.][]{yuan98a,str05,vig03b}
are well characterized by a Gaussian, which justifies the use of this  
estimator in our analysis.} 
For our sample, we obtain $\langle \alpha_{\rm ox} \rangle = -1.65\pm0.05$ 
($-1.62\pm0.05$, for the 5 RQQs only). Our results are compared with those 
of $z\gtrsim 4$ QSOs in Table \ref{tab:aox}. Ideally, we would like to 
compare with $z>4$ QSOs in the same luminosity range 
(-27.0$\gtrsim M_{\rm B}\gtrsim$ -26.3)
\footnote{The magnitude limit used by SSG95 of $M_{\rm B}=-26.0$, 
who used $h_{70}=5/7$, $\Omega_{\rm m}=1$, $\Omega_\Lambda=0$ with 
$\alpha_{\rm o}=-0.5$, corresponds to $M_{\rm B}=-26.47$ and 
-26.55 for our default
cosmology and $\alpha_{\rm o}=-0.79$ at $z=3$ and 5 respectively.}.  
 Unfortunately all but a few of the $z>4$ QSOs observed previously
with X-rays found in literature \citep{bech03,vig01,vig03a,vig05} are more 
luminous ($M_{\rm B}\lesssim -27$) than those in our sample. Keeping 
this limitation
in mind, we compare our result with the mean $\alpha_{\rm ox}$ values 
in Tables 3 and A1 in \citet{vig03c} and Table 3 in \citet{vig05}, 
who used the same methods and assumptions as our work in deriving fluxes 
and $\alpha_{\rm ox}$ values. Their Table A1 includes recalculated 
$\alpha_{\rm ox}$ values of $z>4$ QSOs 
previously observed by {\it Chandra} (including those in \citealt{bech03}) 
using the same method. Their recalculations also took the time-dependent 
low-energy degradation of ACIS-S quantum efficiency into account. 
BALQSOs SDSS 1129-0142 and SDSS 1605-0122 \citep{vig03a} have been 
excluded from the following analysis, because BALQSO 
are known to be X-ray weak due to a heavy absorption. 
Furthermore, we have derived X-ray fluxes of further three $z>4$ QSOs 
from public archival {\it Chandra} ACIS-S data and calculated 
their $\alpha_{\rm ox}$ values in the same way, as summarized 
in Table \ref{tab:arch}.  The 1450 \AA magnitudes of these three have 
been obtained from SSG95 (PC~0910+5625) or the spectra from the SDSS DR3 
database
\footnote{http://www.sdss.org/} after corrections for Galactic 
extinction (the others). These three have been included in our statistical 
analysis.    

\placetable{tab:arch}

\placetable{tab:aox}

\chg{The combined $z>4$ sample is somewhat heterogeneous, as it consists 
of data obtained by a variety of programs, each with different interests 
and strategies.}
In our statistical analysis, we have divided the sample into five groups 
(Group A-E, with Group A being our $z\sim 3$ sample) as shown in 
Table \ref{tab:aox}. 
Figure \ref{fig:aox}(b) shows the scatter diagram of the combined sample 
in the $z$--$M_{\rm B}$ plane with symbols showing the group membership. 
For each group,  we have calculated the mean value 
$\langle \alpha_{\rm ox} \rangle$ and the standard 
deviation of the mean.
The combined sample includes six upper limit $\alpha_{\rm ox}$ values 
(no X-ray detection) and we have used the Kaplan-Meier estimator to calculate 
the mean and standard deviation of the mean using the ASURV 
software\citep{asurv1}, when necessary. Figure \ref{fig:aox}(a) 
and (c) plots $\langle \alpha_{\rm ox} \rangle$ of each group against $z$ and 
$M_{\rm B}$ respectively. 

 Figure \ref{fig:aox}(b) shows a rather \chg{incoherent scatter}, 
reflecting the 
interests of individual original observing programs. As a result, $z$ and 
$M_{\rm B}$ do not show strong monotonic $z$ versus luminosity correlation, 
unlike flux-limited samples. This is fortunate, because if such a correlations 
existed, it would be difficult to separate he redshift and luminosity 
dependences of $\alpha_{\rm ox}$. 

 Figs. \ref{fig:aox}(a) and (c) seems to support the conclusion by \citep{vig03b} 
that the $\alpha_{\rm ox}$ is anti-correlated with ultraviolet luminosity, while 
the correlation with $z$ is weak, if any.  {We made a linear regression
analysis of the dependent parameter $\alpha_{\rm ox}$ against the
two independent parameters, namely  $z$ and $M_{\rm B}$. We also made separate 
one-independent parameter regressions for $z$ or $M_{\rm B}$ versus
the dependent parameter $\alpha_{\rm ox}$.}
 The results the analysis of 
the entire sample of 64 QSOs using the EM method available in ASURV 
\citep{asurv2} are shown below: 
\begin{eqnarray}
\alpha_{\rm ox}=(-1.662\pm.028)+(.042\pm.027)(z-4)\hfill\nonumber\\
                +(.071\pm.019)(M_{\rm B}+27)\hfill\label{eq:reg1}\\ 
               =(-1.737\pm.023)+(.031\pm.029)(z-4)\hfill\label{eq:reg2} \\
               =(-1.648\pm.049)+(.068\pm.019)(M_{\rm B}+27)\hfill\label{eq:reg3}                
\end{eqnarray} 
 The coefficients for the regression analysis with the Buckley-James (B-J) and 
Schmitt's methods (if applicable) that are included in ASURV are consistent
with those from the EM method shown above.  

 From either of Eqs. \ref{eq:reg1} \& \ref{eq:reg3}, we find a 3.5$\sigma$
dependence of $\alpha_{\rm ox}$ to $M_{\rm B}$. The slope coefficeint 
can be translated  into $L_{\rm x}\propto L_{\rm opt}^\beta$ with 
$\beta=0.56\pm 0.12$ (from Eq. \ref{eq:reg3}), that is consistent
with the recent extensive analysis including lower redshift QSOs 
by \citet{str05}.

\begin{inlinefigure}
\centerline{\includegraphics[width=\textwidth]{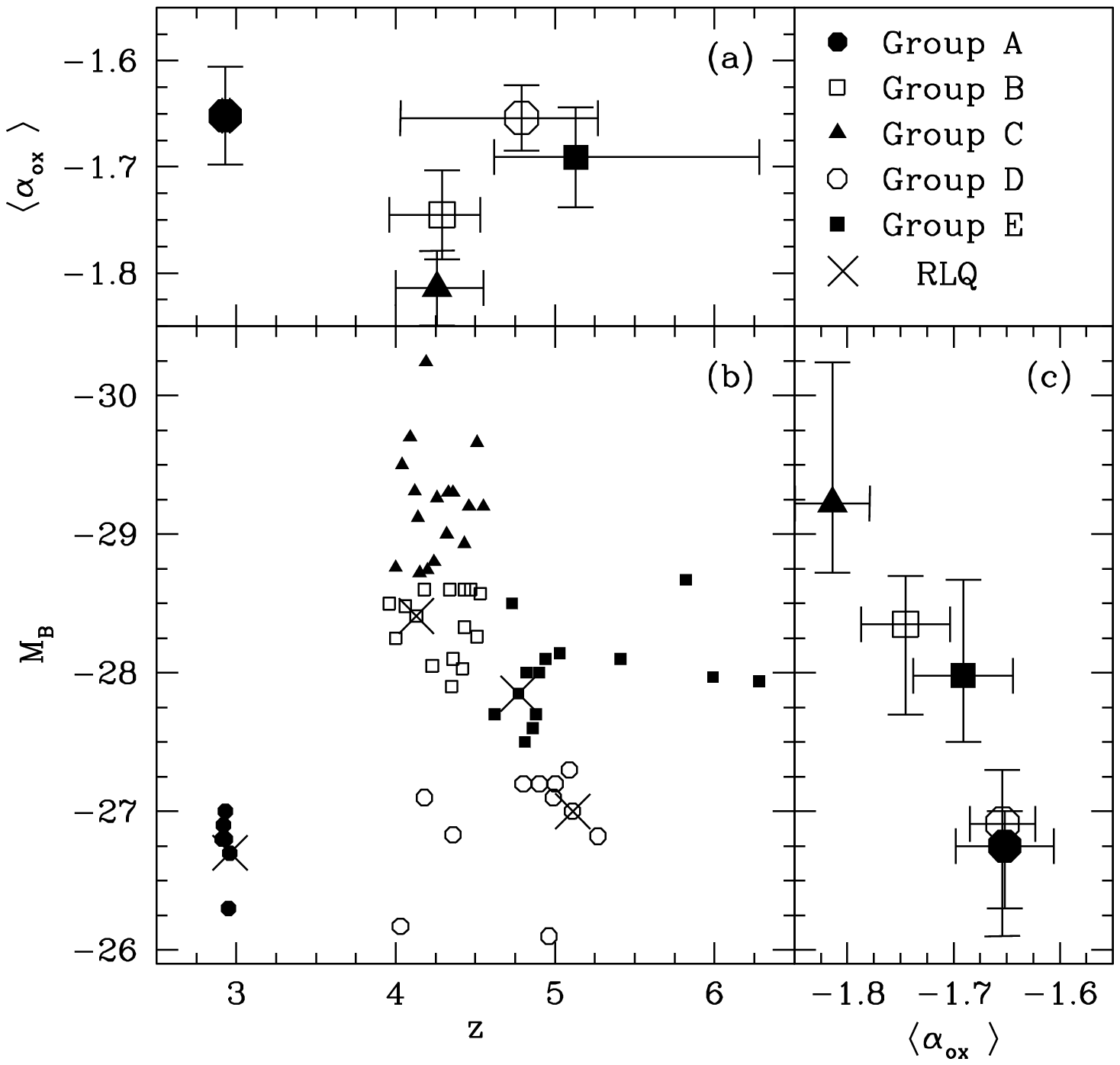}}
\caption{
  Grouping of QSOs from our sample and $z>4$ samples from literature and 
  $\langle \alpha_{\rm ox} \rangle$ of each group plotted against $z$ and 
  $M_{\rm B}$. The QSOs are grouped according to the location in 
 the $z$--$M_{\rm B}$ plane as plotted in panel (b). The groups are 
 discriminated by different symbols as labeled. Radio-detected RLQs are 
 indicated by a large cross.  The $\langle \alpha_{\rm ox} \rangle$ value 
 of each group is plotted against $z$  and $M_{\rm B}$ in panels (a) and 
 (c) respectively. Error bars of  $\langle \alpha_{\rm ox}\rangle$ are the 
 standard deviation of the  mean and those of $z$ and  $M_{\rm B}$ are 
 the range of these values in the group respectively. The data points for 
 the groups are plotted with larger symbols than those for the individual QSOs.
\label{fig:aox}} 
\end{inlinefigure}

\section{Discussion} 

\begin{inlinefigure}
\centerline{\includegraphics[width=\textwidth]{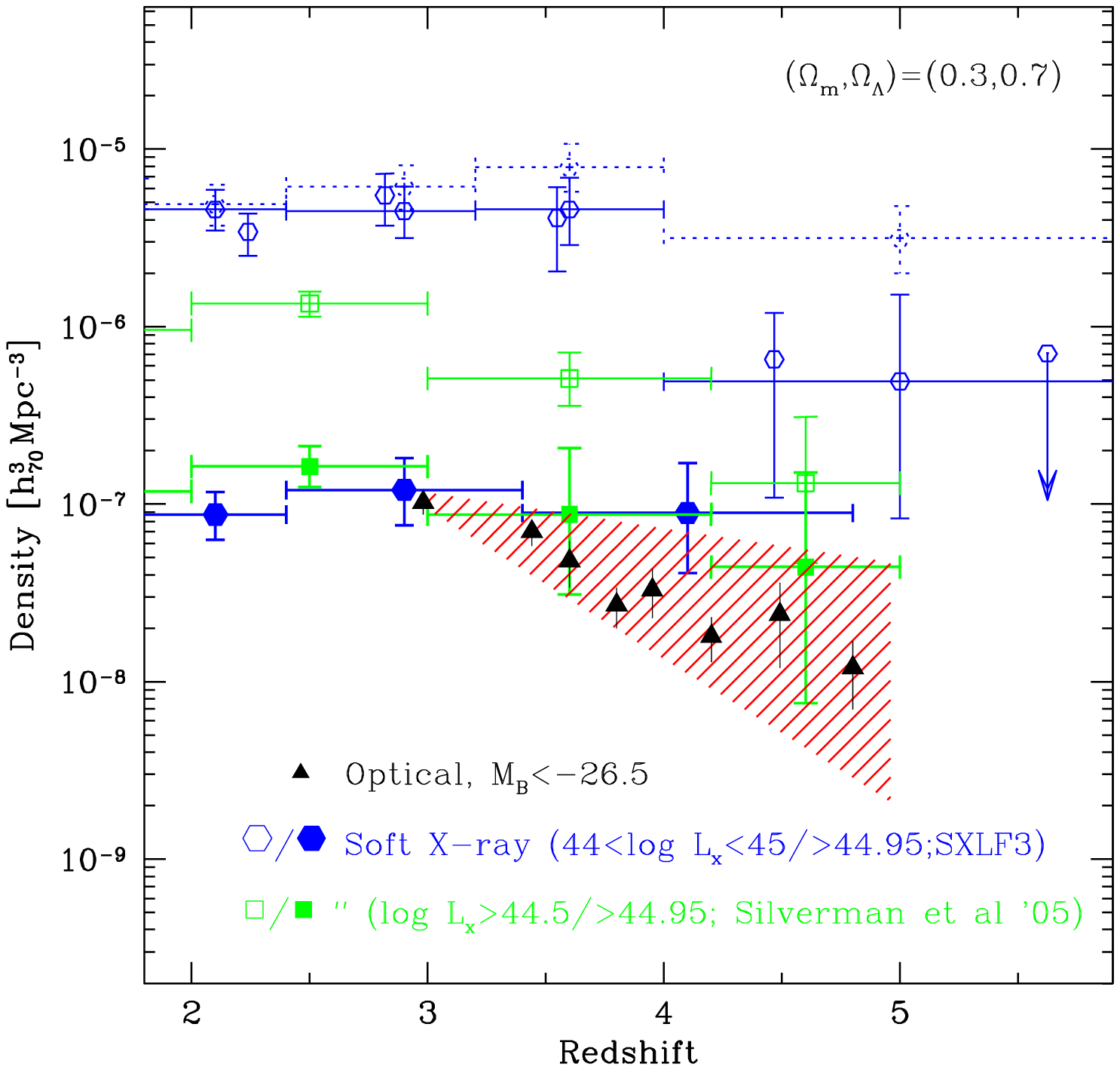}}
\caption{
  The number densities of soft X-ray selected QSOs from samples used by
  SXLF3 (blue hexagons) and \citet{silver05} (green squares) are plotted 
  with those of optically selected QSOs \citep[SSG95;][]{fan01} (triangles), 
  converted to our default cosmology. The X-ray luminosity cut 
  ($\log L_{\rm x}>44.95$) was chosen to give the same number density as the 
  optical curve at $z\sim 2.9$. 
  The hatched area is a 90\% confidence
  range of expected evolution of soft X-ray selected QSOs from the constraint 
  on the systematic $\alpha_{\rm ox}$ change with 
  redshift as determined from the comparison between  groups A \& D. {For 
  reference, the density curves of QSOs  from SXLF3 and 
  \citet{silver05} have been overplotted for the original paper's 
  representative luminosity ranges as labeled and plotted in open symbols.
  The dotted data points from SXLF3 show the absolute maximum number densities
  where all the optically faint unidentified X-ray sources were at the center 
  of the redshift bin.} \label{fig:ndens}} 
\end{inlinefigure}

 Our primary goal of this study is to investigate the redshift dependence of
$\alpha_{\rm ox}$ between $z\sim 3$ and $z\gtrsim 4$ to constrain the
difference of number density behaviors of optically and X-ray selected QSOs 
at $M_{\rm B}\sim -26.5$. In order to achieve this goal, 
we make more careful comparison of groups A (our sample) and D, which are 
well-separated in the redshift range ($\langle z \rangle$ of 2.93 and 4.79 
respectively) and occupy approximately the same luminosity range 
(Fig. \ref{fig:aox}(b)). {Since the difference in the mean $M_{\rm B}$ values
for these two groups are small ($=0.16$), the systematic 
difference of $\alpha_{\rm ox}$ between these groups due 
to optical luminosity dependence is negligible 
($\sim 0.01$ in $\langle \alpha_{\rm ox}\rangle$).} 
The observed $\langle \alpha_{\rm ox} \rangle$ values of these groups are 
essentially the same. We have made Monte-Carlo simulations to find the 
probability distribution of  $\Delta_{\rm DA} \langle \alpha_{\rm ox} \rangle$,
which is the {\em difference} between groups D and A to constrain
the systematic change in $\alpha_{\rm ox}$ with redshift at $z>2.7$ at 
this luminosity range.  In order to make a parent distribution of  
$\alpha_{\rm ox}$ from which the simulations draw objects, we used the all 
QSOs in the current sample except for upper limits.
Also $\alpha_{\rm ox}$ of QSOs in each group have been shifted so that all
groups have the same mean value. The standard deviation of these  
58 $\alpha_{\rm ox}$ values is consistent with those of groups A and D
respectively. In each run, we randomly took 6 QSOs to represent group A and
11 to represent D respectively. The mean of each of the 6 and 11 random 
$\alpha_{\rm ox}$ values has been calculated and the distribution
of the difference of these means 
($\Delta_{\rm DA} \langle \alpha_{\rm ox} \rangle$) for 2000
simulations have been investigated.

 As a result, the range where 90\% of the simulations fall in were
$|\Delta_{\rm DA} \langle \alpha_{\rm ox} \rangle_{\rm DA}|<0.13$.
This limits the systematic difference in $\log (L_{\rm x}/L_{\rm opt})$
between $z\sim 3$ (group A) and $z> 4$ (group D) of $\pm 0.34$. For the 
slope of the luminosity function (LF) of $\gamma_{\rm x}=2$, this 
corresponds to a 
number density difference of 
$|\Delta \log (\rho_{\rm opt}/\rho_{\rm x})|<0.68$, 
where $\rho_{\rm x}$ and $\rho_{\rm opt}$ are the number densities of 
QSOs above  optical luminosity and X-ray luminosities respectively.
Figure \ref{fig:ndens} shows the evolution of comoving number density of
soft X-ray selected QSOs (SXLF3) as a function of redshift plotted with those
of optical QSOs from SSG95 and \cite{fan01} at $z>2.7$ (their original 
values are converted to our default cosmology and $\alpha_{\rm o}$). 
The luminosity limit for the soft X-ray QSOs  is set at 
$\log\; L_{\rm x}>44.95$, where $L_{\rm x}$ is the observed frame 0.5-2 keV in 
${\rm erg\;s^{-1}}$) {corresponding to approximately the $2-8$ keV 
luminosity at the rest frame. Note that this is a rest-frame 0.5-2 keV 
luminosity, under the assumtion of $\Gamma=2.0$ 
power-law spectrum, which is representative of X-ray selected type 1 AGNs
and SXLF3 treats the luminosities as such.} 
We also show the results from
\citep{silver05}'s sample for the same observed 0.5-2 keV luminosity cut.
This luminosity  was selected such that the space density became  
equal to that of optical QSOs at $z\sim 2.7$ in thick lines and 
filled symbols. 
For reference, we show the number density curves from SXLF3 and
\citep{silver05} (open symbols with thin lines), in lower 
luminosity ranges (see labels), which are representative of the 
respective samples. These show declines in $z\gtrsim 3$.  The large error 
bars in the $\log\; L_{\rm x}>44.95$ data at $z>2.7$ show that the QSOs 
in this regime 
is still underrepresented by current X-ray surveys.   
The limits of the number density curve of X-ray selected sample are shown 
in the shaded area in  Fig. \ref{fig:ndens}, corresponding to 
$|\Delta \langle \alpha_{\rm ox} \rangle_{\rm DA}|<0.13$.

  A limitation of the above investigation is the 
uncertainties in the effects of the sample selection and variability.
The X-ray and optical luminosities have been measured in different
epochs. Thus the variability of AGNs have a net effect of increasing 
the variance of the $\alpha_{\rm ox}$ distribution. Our underlying 
assumption is that the variability does not cause a net systematic
difference in its effect on the mean $\alpha_{\rm ox}$ between sample
A and sample D,\chg{separeted in redshift, but not in luminosity.} 
Both are optically selected samples and
$\langle \alpha_{\rm ox} \rangle$ should be biased towards larger
(more optically luminous) values than the ``true'' 
$\langle \alpha_{\rm ox} \rangle$ (i.e. $\alpha_{\rm ox}$ of
time-averaged mean optical and X-ray luminosities), because optical 
selection is more likely to pick up the AGN when it is more 
optically-bright, while the X-ray followup of the same object 
typically gives average X-ray luminosity of the source. 
As long as both are selected in the optical and followed up by
X-ray, that the slope of the LFs 
are the same at both redshifts, and that there is no systematic 
difference in the variability amplitudes of AGNs with redshift, the effect
of this \chg{``variability''} bias should be the equal between sample A and 
sample D. \chg{Thus we do not expect that the variabilty bias plays a
major role in our analysis on the redshift dependence of 
$\langle \alpha_{\rm ox} \rangle$.}      

 From a combination of our sample at $z\sim 3$ and $z>4$ QSOs observed by 
{\it Chandra}, we have confirmed the apparent dependence of $\alpha_{\rm ox}$ 
on optical luminosity. This result, however, should be
treated with caution, \chg{because our sample is optically selected
and subject to the variability effect of preferentially  picking 
up optically brighter phase as described above.} 
An X-ray selected sample covering a much larger regime in $z$-$L$ 
space rather showed $L_{\rm x}\propto L_{\rm opt}$ \citep{has05}. This 
is, however subject to a similar effect working in the opposite sense.
At this limited regime, on the other hand, there is a hint that 
\chg{our presently
determined} dependence might reflect the true behavior of the shift of 
$\alpha_{\rm ox}$ with QSO power.
 Expressing the LF as $d\Phi/d \log\, L\propto L^{-\gamma}$
at the luminosities of interest, the soft X-ray LF has $\gamma_{\rm X}=2.2\pm 0.3$ 
and (90\% error) in $2.7<z<4.8$ (from the same sample as SXLF3). This can 
be compared with the optical LF of $\gamma_{\rm opt}=1.6\pm 0.2$ (1$\sigma$ error) 
by \citet{fan01} or $\gamma_{\rm opt}=1.87$ by \citet{ssg95}. The trend that the 
optical LF has a flatter slope than the soft X-ray counterpart is consistent 
with the relation $\gamma_{\rm opt}=\beta\gamma_{\rm x}$ within errors. 
This is also consistent with the comparison between X-ray and 
optically-selected AGN luminosity functions by \citet{ueda03} 
(see their Fig. 20), where a conversion of their hard X-ray LF
to the optical band assuming $L_{\rm x}\propto L_{\rm opt}^{0.70}$ 
gave a  good match to an observed optical QSO LF at high 
luminosities. However, our most recent comparison of the SXLF
\citep{sxlf3} (high luminosity end) and optical QSO LF by 
\citet{croom04} at $z<2.1$ is more consistent with 
$L_{\rm x}\propto L_{\rm opt}$, thus a more study is needed to
investigate the relationship between direct comparison of $L_{\rm x}$  
and $L_{\rm opt}$ and the conversion between the X-ray and optically
selected QSO LFs.

\section{Summary}
 
{We have made {\it Chandra} ACIS-S observations of six QSOs at $z\sim 3$,
which marks the peak of luminous QSO number density. These observatios 
fill a redshift gap in the X-ray coverage of luminous QSOs. We found an average 
photon index of $\langle \Gamma \rangle=1.9\pm 0.3$ from the stacked spectrum
and we also found $\langle \alpha_{\rm ox} \rangle = -1.65\pm .05$. The
$\langle \alpha_{\rm ox} \rangle$ value is essentially the same as those
at $z>4$ in the similar UV luminosity range and thus we have found no 
systematic shift of X-ray to UV luminosity ratios with redshuft above $z=3$. 
The density curves of $M_{\rm B}<-26.5$ optically selected QSOs
and $\log L_{\rm x}>44.92$ soft X-ray selected QSOs, giving the same 
densities at $z\sim 2.7$, are statistically consistent with each other
within our limit of the systematic $\langle \alpha_{\rm ox} \rangle$
shift at $z>3$. We note that this regime is still underrepresented by X-ray 
surveys. Large-area moderarely-deep X-ray surveys are needed to trace 
the rise of number density of the most luminous QSOs at $z>3$ in X-rays.}      

\acknowledgments
 This work has been supported by Chandra General Observer Award 
GO3-4153X, NASA LTSA Grant NAG5-10875 (TM) and NSF grant AST03-07582 (DPS). 
We thank John Silverman for calculating the space density curves for our
luminosity cuts and cosmological parameter choice. 
  



\begin{deluxetable}{ccrccr}
\footnotesize
\tablecaption{Log of Chandra Observations and Sample Properties\label{tab:log}}
\tablewidth{0pt}
\tablehead{
\colhead{Name} & 
\colhead{Obsid/Date} & 
\colhead{Expo.} & 
\colhead{z} & 
\colhead{$AB_{\rm 1450}$\tablenotemark{a}} & 
\colhead{$R_{\rm L}$} \\
&                  & \colhead{(ks)}\\
}

\startdata
 PC 0041+0215 &  4150/2003 Sep 01 &  9.2 & 2.93 & 19.5 &  $<1.5$\\
 PC 0947+5628 &  4151/2003 Jan 25 &  9.0 & 2.91 & 19.5 &  $<1.0$\\
 PC 1000+4751 &  4152/2002 Dec 18 & 13.9 & 2.95 & 20.0 &  $<1.2$\\
 PC 1015+4752 &  4153/2003 Jan 01 &  8.1 & 2.92 & 19.4 &  $<0.9$\\
 PC 1035+4747 &  4154/2003 Mar 16 & 10.0 & 2.96 & 19.6 &  2.3\\ 
 PC 1447+4750 &  4155/2003 Jul 28 &  7.0 & 2.93 & 19.3 &  $<0.9$\\
\enddata
\tablenotetext{a}{Some authors prefer to use the notation ``$AB_{1450(1+z)}$''.}
\end{deluxetable}

\begin{deluxetable}{cccccccccc}
\footnotesize
\rotate
\tablecaption{X-ray Spectral Analysis Results and Derived Quantities\label{tab:spec}}
\tablewidth{0pt}
\tablehead{
\colhead{Name} & 
\colhead{$z$} & 
\colhead{$N_{\rm H,20}^{\rm G}$\tablenotemark{a}}&
\colhead{$\Gamma$} & 
\colhead{$S_{\rm x,14}^{\rm int}$\tablenotemark{b}}& 
\colhead{$N_{\rm H,20}^{\rm z}$\tablenotemark{a}} & 
\colhead{Scts} & 
\colhead{$\log\;L_x$\tablenotemark{c}} & 
\colhead{$M_{\rm B}$} &
\colhead{$\alpha_{\rm ox}$}  
}
\startdata
 PC 0041+0215&2.93& 2.8&2.4(1.9;2.9) & 1.4(1.0;1.8)&0&36&45.0&$-26.8$&$-1.56$\\
 PC 0947+5628&2.91& 1.0&1.5(0.7;2.3) & .31(.17;.52)&0&10&44.4&$-26.8$&$-1.78$\\
 PC 1000+4751&2.95& 0.9&1.5(1.2;1.9) & 1.2(1.0;1.5)&0&60&45.0&$-26.3$&$-1.48$\\
 PC 1015+4752&2.92& 1.0&2.4(1.8;3.0) & .94(.65;1.2)&0&24&44.9&$-26.9$&$-1.64$\\
 PC 1035+4747&2.96& 1.3&0.6($-0.3$;1.4)&.17(.08;.34)&0& 9&44.3&$-26.7$&$-1.80$\\
 PC 1447+4750&2.93& 2.0&2.0(1.4;2.6) & 1.0(.64;1.4)&0&21&44.9&$-27.0$&$-1.65$\\ 
$\langle$6 QSOs$\rangle$  &2.93& ...&1.9(1.7;2.1) &     ...  & $<90$ & &160& ...\\
\enddata

\tablecomments{The 90\% confidence ranges and upper limits for 
one interesting parameter are shown for free parameters. Spectral 
parameters without a confidence range were fixed during the fit.}
\tablenotetext{a}{In units of $10^{20}{\rm cm^{-2}}$.}
\tablenotetext{b}{X-ray flux before Galactic absorption in 0.5-2 keV 
  (observer's frame), in  units of $10^{-14}{\rm erg\;s^{-1}\;cm^{-2}}$.}
\tablenotetext{c}{Base-10 logarithm of X-ray luminosity at the rest-frame 2-10 keV band in 
 units of $h_{70}^{-2}{\rm \;erg\;s^{-1}}$.}
\end{deluxetable}

\begin{deluxetable}{ccrcccccrcc}
\footnotesize
\rotate
\tablecaption{Additional {\it Chandra} Archival Data Analysis\label{tab:arch}}
\tablewidth{0pt}
\tablehead{
\colhead{Name} & 
\colhead{z} &
\colhead{Obsid/Date} & 
\colhead{Expo.}  &
\colhead{$N_{\rm H,20}^{\rm G}$\tablenotemark{a}}    &
\colhead{$S_{\rm x,14}^{\rm int}$\tablenotemark{a}\tablenotemark{b}} & 
\colhead{$AB_{\rm 1450}$} & 
\colhead{$R_{\rm L}$} & $M_{\rm B}$ & $\alpha_{\rm ox}$ \\
&                  & \colhead{(ks)}\\
}
\startdata
 PC~0910+5625             & 4.04 & 4821/2004 Mar 28 & 23. & 2.9 & 
               1.7(1.0;2.7) & 20.7 & $<1.5$&-26.2&-1.67\\
 SDSS~J235718.36+004350.3 & 4.36 & 4827/2003 Nov 26 & 12. & 3.3 & 
               5.0(3.3;7.2) & 20.2 & $<1.2$&-26.9&-1.56\\
 SDSS~J144428.67-012344.1 & 4.17 & 4826/2004 Jan 8  & 10. & 4.0 & 
               1.8(0.7;3.4) & 19.8 & $<1.2$&-27.1&-1.79\\
\enddata
\tablenotetext{a}{See notes to Table~\ref{tab:spec} for units.}
\tablenotetext{b}{90\% confidence ranges are shown.}
\end{deluxetable}

\begin{deluxetable}{cccccrr}
\footnotesize
\tablecaption{Comparison of $\alpha_{\rm ox}$\label{tab:aox}}
\tablewidth{0pt}
\tablehead{
\colhead{group} &
\colhead{$z_{\rm min},z_{\rm max}$} & 
\colhead{$\langle z \rangle$} & 
\colhead{$M_{\rm B,min},M_{\rm B,max}$} &
\colhead{$\langle M_{\rm B} \rangle$}& 
\colhead{$N_{\rm Q}$}& 
\colhead{$\langle\alpha_{\rm ox}\rangle$\tablenotemark{a}}
}
\startdata
 A &2.9,3.0 & 2.93 &$-27.0,-26.3$ &$-26.75$& 6& $-1.652\pm$.046 \\
 B &3.5,4.6 & 4.29 &$-28.7,-27.7$ &$-28.35$&15& $-1.745\pm$.042 \\
 C &3.5,4.6 & 4.26 &$-30.3,-28.7$ &$-29.22$&17& $-1.814\pm$.035 \\
 D &4.0,5.3 & 4.79 &$-27.5,-26.0$ &$-26.91$&11& $-1.654\pm$.031 \\
 E &4.6,6.4 & 5.13 &$-28.8,-27.5$ &$-27.98$&14& $-1.691\pm$.047 \\
\enddata
\tablenotetext{a}{The errors are the standard deviation of the mean.}
\end{deluxetable}






\begin{thebibliography}{}
\bibitem[Barger et al.(2005)]{barger05} Barger, A.~J., Cowie, L.~L., 
Mushotzky, R.~F., Yang, Y., Wang, W.-H., Steffen, A.~T., \& Capak, 
P.\ 2005, \aj, 129, 578 
\bibitem[Bechtold et al.(2003)]{bech03} Bechtold, J., et al.\ 
2003, \apj, 588, 119
\bibitem[Becker, White, \& Helfand(1995)]{first} Becker, 
R.~H., White, R.~L., \& Helfand, D.~J.\ 1995, \apj, 450, 559
\bibitem[Boyle et al.(1988)]{boy88} Boyle, B.~J., Shanks, T., 
\& Peterson, B.~A.\ 1988, \mnras, 235, 935 
\bibitem[Cash(1979)]{cash} Cash, W. 1979, \apj, 228, 939
\bibitem[Condon et al.(1998)]{nvss} Condon, J.~J., Cotton, 
W.~D., Greisen, E.~W., Yin, Q.~F., Perley, R.~A., Taylor, G.~B., \& 
Broderick, J.~J.\ 1998, \aj, 115, 1693
\bibitem[Cowie et al.(2003)]{cowie03} Cowie, L.~L., Barger, 
A.~J., Bautz, M.~W., Brandt, W.~N., \& Garmire, G.~P.\ 2003, \apjl, 584, 
L57 
\bibitem[Croom et al.(2004)]{croom04} Croom, S.~M., Smith, 
R.~J., Boyle, B.~J., Shanks, T., Miller, L., Outram, P.~J., \& Loaring, 
N.~S.\ 2004, \mnras, 349, 1397 
\bibitem[Dickey \& Lockman(1990)]{nh} Dickey, J.~M.~\& 
Lockman, F.~J.\ 1990, \araa, 28, 215
\bibitem[Isobe, Feigelson, \& Nelson(1986)]{asurv2} Isobe, T., 
Feigelson, E.~D., \& Nelson, P.~I.\ 1986, \apj, 306, 490 
\bibitem[Fan et al.(2001)]{fan01} Fan, X., et al.\ 2001, \aj, 
121, 54
\bibitem[Feigelson \& Nelson(1985)]{asurv1} Feigelson, 
E.~D.~\& Nelson, P.~I.\ 1985, \apj, 293, 192
\bibitem[Fiore et al.(2003)]{fiore03} Fiore, F., et al.\ 2003, 
\aap, 409, 79 
\bibitem[Garmire et al.(2003)]{acis} Garmire, G.~P., Bautz, 
M.~W., Ford, P.~G., Nousek, J.~A., \& Ricker, G.~R.\ 2003, 
\procspie, 4851, 28 
\bibitem[Hasinger(2005)]{has05} Hasinger, G. 2005 in ``Growing Black Holes"
 eds A. Merloni, S. Nayakshin and R. Sunyaev (Heidelberg:Springer), p418
\bibitem[Hasinger,Miyaji\& Schmidt(2005)]{sxlf3} Hasinger, G., Miyaji, T.,
 Schmidt, M. 2005, \aap, 441, 417 
\bibitem[Miyaji, Hasinger, \& Schmidt(2000)]{xlf1}
      Miyaji, T., Hasinger, G. \& Schmidt, M. 2000, \aap, 353, 25 (SXLF1)
\bibitem[Osmer(1982)]{osm82} Osmer, P.~S.\ 1982, \apj, 253, 28 
\bibitem[Schneider, Schmidt, \& Gunn(1994)]{ssg94} Schneider, 
  D.~P., Schmidt, M., \& Gunn, J.~E.\ 1994, \aj, 107, 1245
\bibitem[Schneider et al.(2001)]{dps01} Schneider, D.~P., et 
al.\ 2001, \aj, 121, 1232 
\bibitem[Schmidt, Schneider, \& Gunn(1995)]{ssg95} Schmidt, 
  M., Schneider, D.~P., \& Gunn, J.~E.\ 1995, \aj, 110, 68 (SSG95)
\bibitem[Shaver et al.(1996)]{shav96} Shaver, P.~A., Wall, 
J.~V., Kellermann, K.~I., Jackson, C.~A., \& Hawkins, M.~R.~S.\ 1996, \nat, 
384, 439
\bibitem[Silverman et al.(2005)]{silver05} Silverman, J.~D., et 
al.\ 2005, \apj, 624, 630 
\bibitem[Strateva et al.(2005)]{str05} Strateva, I.~V., 
Brandt, W.~N., Schneider, D.~P., Vanden Berk, D.~G., \& Vignali, C.\ 2005, 
\aj, 130, 387 
\bibitem[Ueda et al.(2003)]{ueda03} Ueda, Y., Akiyama, M., 
  Ohta, K., \& Miyaji, T.\ 2003, \apj, 598, 886 
\bibitem[Vignali et al.(2001)]{vig01} Vignali, C., Brandt, 
W.~N., Fan, X., Gunn, J.~E., Kaspi, S., Schneider, D.~P., \& Strauss, 
M.~A.\ 2001, \aj, 122, 2143 
\bibitem[Vignali et al.(2003a)]{vig03a} Vignali, C., Brandt, 
W.~N., Schneider, D.~P., Garmire, G.~P., \& Kaspi, S.\ 2003a, \aj, 125, 418 
\bibitem[Vignali, Brandt, \& Schneider(2003b)]{vig03b} Vignali, C., 
  Brandt, W.~N., \& Schneider, D.~P.\ 2003b, \aj, 125, 433 
\bibitem[Vignali et al.(2003c)]{vig03c} Vignali, C., et al.\ 
2003c, \aj, 125, 2876
\bibitem[Vignali et al.(2005)]{vig05} Vignali, C., Brandt, 
W.~N., Schneider, D.~P., \& Kaspi, S.\ 2005, \aj, 129, 2519 
\bibitem[Wall et al.(2005)]{wall05} Wall, J.~V., Jackson, 
C.~A., Shaver, P.~A., Hook, I.~M., \& Kellermann, K.~I.\ 2005, \aap, 434, 
133
\bibitem[Warren, Hewett, \& Osmer(1994)]{who94} Warren, 
S.~J., Hewett, P.~C., \& Osmer, P.~S.\ 1994, \apj, 421, 412
\bibitem[Wilkes(2000)]{wilkes} Wilkes, B. J. 2000, in ``Allen's Astrophysical
  Quantities, Fourth Edition'' ed. A. N. Cox (New York:Springer), Chap. 24
\bibitem[York et al.(2000)]{sdss} York, D.~G., et al.\ 2000, 
\aj, 120, 1579
\bibitem[Yuan et al.(1998a)]{yuan98a} Yuan, W., Brinkmann, W., 
Siebert, J., \& Voges, W.\ 1998, \aap, 330, 108 
 \bibitem[Yuan et al.(1998b)]{yuan98b} Yuan, W., Siebert, J., \& 
Brinkmann, W.\ 1998, \aap, 334, 498 
\end{thebibliography}
\end{document}